\documentclass[11pt,journal,twoside,onecolumn,draftcls]{IEEEtran}

\usepackage{graphicx,subfigure}
\usepackage{amsfonts,amsmath,amssymb}
\usepackage{epic,eepic,eepicemu}
\usepackage{epsf}
\usepackage{epsfig}
\usepackage{graphics}
\usepackage{psfrag}

\makeatletter
\def\@cite#1#2{[\if@tempswa #2 \fi #1]}
\makeatother

\newcommand{\myparagraph}[1]{{\bf{#1}}}

\newcommand{\widgraph}[2]{\includegraphics[keepaspectratio,width=#1]{#2}}

\newcommand{\beq}{\begin{equation}}
\newcommand{\eeq}{\end{equation}}
\newcommand{\bea}{\begin{eqnarray}}
\newcommand{\eea}{\end{eqnarray}}

\def\qed{\quad \vrule height6.5pt width6pt depth0pt} %Nifty end proof sign
\def\qed{\quad \vrule height6.5pt width6pt depth0pt} %Nifty end proof sign
\newtheorem{atheorem}{Theorem}
\newtheorem{lemma}{Lemma}
\newtheorem{definition}{Definition}
\newtheorem{proposition}{Proposition}
\newtheorem{corollary}{Corollary}

\long\def\symbolfootnote[#1]#2{\begingroup
\def\thefootnote{\fnsymbol{footnote}}\footnote[#1]{#2}\endgroup}

\newcommand{\const}{\ensuremath{C}}
\newcommand{\nbit}{\ensuremath{n}}
\newcommand{\mcheck}{\ensuremath{m}}
\newcommand{\Neigh}{\ensuremath{N}}
\newcommand{\defn}{\ensuremath{: \, =}}

\newcommand{\actset}{\ensuremath{\mathbb{A}}}
\newcommand{\cw}{\ensuremath{x^{\operatorname{cw}}}}
\newcommand{\pcw}{\ensuremath{x^{\operatorname{pc}}}}
\newcommand{\rate}{\ensuremath{R}}
\newcommand{\cdeg}{\ensuremath{d_c}}
\newcommand{\vdeg}{\ensuremath{d_v}}
\newcommand{\vertex}{\ensuremath{V}}

\newcommand{\mybeginproof}{\noindent \emph{Proof: $\;$}}
\newcommand{\myendproof}{\hfill \qed}

\newcommand{\vertfrac}{\ensuremath{\vertex_{\operatorname{frac}}}}
\newcommand{\checkfrac}{\ensuremath{C_{\operatorname{frac}}}}

\newcommand{\Forbid}{\ensuremath{\mathbb{F}}}
\newcommand{\Boxin}{\ensuremath{\mathbb{B}}}

\newcommand{\checknum}{\ensuremath{m}}
\newcommand{\bitnum}{\ensuremath{n}}
\newcommand{\plaincw}{\ensuremath{x}}
\newcommand{\CheckMat}{\ensuremath{H}}
\newcommand{\Code}{\ensuremath{\mathbb{C}}}

\newcommand{\graph}{\ensuremath{G}}

\newcommand{\RelPoly}{\ensuremath{\mathcal{P}}}

\newcommand{\xml}{\widehat{x}^{\operatorname{ML}}}

\long\def\comment#1{}

\begin{document}

\title{\huge{Guessing Facets: Polytope Structure and \\
Improved LP Decoding}}


\author{Alexandros G. Dimakis, Amin A. Gohari, and Martin
J. Wainwright$^{1,2}$ \\ $^1$ Department of Electrical Engineering and
Computer Science \\ $^2$ Department of Statistics \\ University of
California, Berkeley\\ \texttt{\small
$\{$adim,aminzade,wainwrig$\}$@eecs.berkeley.edu} \\ }

% avoiding spaces at the end of the author lines is not a problem with
% conference papers because we don't use \thanks or \IEEEmembership

\maketitle

% NO PREPRINT LABEL HERE
\comment{
\begin{center}
%% PREPRINT LABEL:
\vspace*{-2.5in}
\vbox to 2.5in{\footnotesize {\tt
 \begin{tabular}[t]{c}
    Appeared in: \\
    International Symposium on Information Theory \\
    Seattle, WA;  July 2006
  \end{tabular} \vfil}}
\end{center}
}
% END COMMENT

\begin{abstract}
In this paper we investigate the structure of the fundamental polytope used in
the Linear Programming decoding introduced by Feldman, Karger and
Wainwright.  We begin by showing that for expander codes, every
fractional pseudocodeword always has at least a constant fraction of
non-integral bits.  We then prove that for expander codes, the active
set of any fractional pseudocodeword is smaller by a constant fraction
than the active set of any codeword. We further exploit these geometrical properties
to devise an improved decoding algorithm with the same complexity order as LP decoding
that provably performs better, for any blocklength. It proceeds by
guessing facets of the polytope, and then resolving the linear program
on these facets.  While the LP decoder succeeds only if the ML
codeword has the highest likelihood over all pseudocodewords, we prove
that the proposed algorithm, when applied to suitable expander codes,
succeeds unless there exist a certain number of pseudocodewords, all
adjacent to the ML codeword on the LP decoding polytope, and with
higher likelihood than the ML codeword.  We then describe an extended
algorithm, still with polynomial complexity, that succeeds as long as
there are at most polynomially many pseudocodewords above the ML codeword.
\end{abstract}

\footnote{This work was presented in part at the Interational
Symposium on Information Theory, Seattle, WA, July 2006.}

\section{Introduction}

Low-density parity check (LDPC) codes are a class of graphical codes,
originally introduced by Gallager~\cite{Gallager63}, that come very
close to capacity for large blocklengths even when decoded with the
sub-optimal sum-product algorithm.  The standard techniques for
analyzing the sum-product algorithm, including density
evolution~\cite{Richardson01a} and EXIT charts~\cite{Ashikhmin04}, are
asymptotic in nature.  Many applications, however, require the use of
intermediate blocklengths, for which methods of an asymptotic nature
may not be suitable for explaining or predicting the behavior of the
decoding algorithms.  Feldman, Karger and Wainwright~\cite{Feldman05}
introduced the LP decoding method, which is based on solving a
linear-programming relaxation of the integer program corresponding to
the maximum likelihood (ML) decoding problem.  In practical terms, the
performance of LP decoding is roughly comparable to min-sum decoding
and slightly inferior to sum-product decoding.  In contrast to
message-passing decoding, however, the LP decoder either concedes
failure on a problem, or returns a codeword along with a guarantee
that it is the ML codeword, thereby eliminating any undetected
decoding errors.  On the conceptual level, the correctness of LP
decoding reduces to geometric questions about cost vectors and
polytope structure, so that the method is well-suited to questions of
finite-analysis.  Indeed, all of the analysis in this paper applies to
finite-length codes.  \\

\subsection{Background and previous work}

Feldman et al. first introduced and studied the basic idea of LP
decoding for turbo and low-density parity check
codes~\cite{Feldman03,Feldman05}.  There are various connections to
message-passing~\cite{FelKarWai02,Wainwright02aller}, including links
between the reweighted max-product algorithm and dual LP
relaxations~\cite{Wainwright02aller,WaiJaaWil05b}, and the standard
max-product algorithm and graph covers~\cite{KoeVon03}.  For the
binary symmetric channels and suitable expander codes, it has been
shown that LP decoding can correct a linear fraction
of random~\cite{DasDimKarWai06} or adversarial~\cite{Feldman07} bit-flipping errors.
  Koetter and Vontobel~\cite{kv_bethe,KoeVon03} established
bounds on the pseudo-weight for the additive white Gaussian noise
(AWGN) channel, showing that it grows only sublinearly for regular
codes, and hence that the error probability of LP decoding cannot
decay exponentially for the AWGN channel. Subsequent
work~\cite{FelKoeVon05} exploited the constant fraction
guarantee~\cite{Feldman07} to show that LP decoding error decays
exponentially for Gaussian channels if the likelihoods are suitably
truncated.  Other researchers have studied efficient algorithms for
solving the LP relaxation, including the reweighted max-product
algorithm~\cite{WaiJaaWil05b}, other forms of iterative dual
algorithms~\cite{VonKoe06b}, and adaptive procedures~\cite{TagSie06}.
As with the work described here, a related line of work has studied
various improvements to either standard iterative
decoding~\cite{Fossorier01,PishroFekri} or to LP decoding via
nonlinear extensions~\cite{YanFelWan06} or loop
corrections~\cite{CheChe06}.

\subsection{Our contributions}

The LP decoder operates by solving a linear program over a polytope
$\mathcal{P}$ which constitutes a relaxation of the original
combinatorial codeword space. The polytope $\mathcal{P}$ (referred in the literature as
\emph{relaxed polytope} or \emph{fundamental polytope}) has two types
of vertices: \emph{integral vertices} with $0-1$ components
corresponding to codewords, and \emph{fractional vertices} that
correspond to pseudocodewords.  The first contribution of this paper
is to characterize several geometric properties of this relaxed
polytope for suitable classes of expander codes. For a given
(fractional) pseudocodeword, we define the fractional support as the
subset of coordinates that are non-integral.  For general codes,
there may exist pseudocodewords with very small fractional supports.
Our first result is to show that that for suitable classes of expander
codes, the fractional support always scales linearly in the
blocklength.  In conjunction with known results on the AWGN
pseudoweight~\cite{kv_bethe,KoeVon03}, this fact implies that the size
of the minimal non-zero entry in these pseudocodewords must be
vanishing at a rate faster than inverse blocklength.  In addition, we
show that the relaxed polytope $\mathcal{P}$ has the property that
many more (a constant fraction of the blocklength) facets are adjacent to integral vertices relative to fractional
ones.

Motivated by this geometric intuition, we propose an improved LP
decoding algorithm that eliminates fractional pseudocodewords by
guessing facets of $\mathcal{P}$, and then decodes by re-solving the
optimization problem on these facets. We also provide some theoretical
performance guarantees on this improved solver: in particular, for
suitable expander codes we prove that it always succeeds as long as
there are at most some constant number of pseudocodewords with higher
likelihood than the ML codeword.  Despite the relative
conservativeness of this guarantee, our experimental results show
significant performance improvements, particularly at high SNR, for
small and moderate blocklengths.  In addition, we analyze another type
of randomized facet-guessing, still with polynomial complexity, and
prove that it succeeds as long as there at most a polynomial number of
pseudocodewords, all adjacent to the ML codeword and with higher
likelihood.  Although previous work~\cite{kv_bethe,KoeVon03} shows
that for the AWGN channel, there do exist pseudocodewords with
sublinear pseudoweight, this improved algorithm can fail only if there
exist a super-polynomial number of such pseudocodewords.  Therefore,
our paper raises the interesting question as to the number of
pseudocodewords with sublinear pseudoweight.

%From a practical point of view, we propose a family of algorithms that generalize bit-guessing.

The remainder of this paper is organized as follows.  In
Section~\ref{SecBackground}, we provide background on low-density
parity check codes and linear programming relaxations for decoding.
Section~\ref{SecStructural} presents some results on the structure of
the LP decoding polytope for suitable classes of expander codes.  In
Section~\ref{SecImproved} and~\ref{SecHigher}, we describe and analyze
improved versions of LP decoding that leverage these structural
properties.  We conclude with a discussion in
Section~\ref{SecDiscussion}.

\section{Background}
\label{SecBackground}

In this section, we provide basic background on binary linear codes,
factor graphs, and decoding based on linear programming.

\subsection{Low-density parity check codes and factor graphs}

A binary linear code of blocklength $\bitnum$ consists of a linear
subspace, where arithmetic is performed modulo two, of the set of all
binary sequences $x \in \{0,1 \}^\bitnum$.  A code of rate $\rate = 1
- \frac{\checknum}{\bitnum}$ can be specified by a parity check matrix
$\CheckMat \in \{0,1\}^{\checknum \times \bitnum}$: in particular, the
code $\Code$ consists of all vectors $\plaincw \in \{0,1\}^\bitnum$
that satisfy $\CheckMat \plaincw = 0$ in modulo two arithmetic.  Of
interest in this paper are \emph{low-density parity check} (LDPC)
codes~\cite{Gallager63}, meaning that the number of ones in each row
and column of the parity check matrix remains bounded independently of
the blocklength.

A convenient representation of any LDPC code is in terms of its factor
graph~\cite{Tanner80,Frank01}, a bipartite graph consisting of a set
of variable nodes $V = \{1, \ldots, \bitnum \}$ corresponding to the
columns of $\CheckMat$, and a set of factor nodes $C = \{1, \ldots,
\checknum \}$ corresponding to the rows of $\CheckMat$.  The factor
graph has an edge $(i,a)$ between bit $i$ and check $a$ if and only if
$\CheckMat_{ai}$, so that check $a$ imposes a constraint on bit $i$.

\subsection{Decoding via linear programming}

The problem of maximum likelihood (ML) decoding is to determine the
most likely codeword on the basis of an observation from a noisy
channel.  For a binary linear code, the ML decoding problem can be
formulated as an integer program of the form
\begin{eqnarray}
\label{EqnMLIP}
\xml & \defn & \arg \max_{x \in \Code} \sum_{i=1}^\bitnum \gamma_i x_i,
\end{eqnarray}
where $\gamma_i$ is a channel-dependent quantity.  As a concrete
illustration, if the all-zeroes codeword were transmitted over the
binary symmetric channel with cross-over $\epsilon \in (0,
\frac{1}{2})$, then $\gamma_i = -1$ with probability $1-\epsilon$, and
$\gamma_i = +1$ with probability $\epsilon$.  In the absence of
additional structure, the ML decoding problem~\eqref{EqnMLIP} requires
an exhaustive search over an exponentially large space, and is known
to be computationally difficult~\cite{Berlekamp78}.  It is
straightforward to convert the ML decoding problem into a \emph{linear
program} over the convex hull of all codewords, a polytope known as
the codeword polytope~\cite{Feldman05}.  Unfortunately, for a general
binary linear code, the number of constraints (or facets) required to
characterize the codeword polytope grows in a non-polynomial manner
with blocklength.  Even more strongly, the existence of a
polynomial-time separation oracle for the codeword polytope of a
general linear code is very unlikely since ML decoding for arbitrary
linear codes is NP-hard~\cite{Berlekamp78}. \\

The technique of LP decoding~\cite{Feldman05} is based on solving a
linear program over a relaxation of the codeword polytope that can be
characterized with a polynomial number of inequalities.  The standard
approach to date is based on a polytope $\mathcal{P}$ obtained by enforcing a set of local constraints associated with each bit.  This
polytope is referred to as a tree-based relaxation, since it is
guaranteed to be exact for any tree-structured factor
graph~\cite{WaiJaaWil05b}.

In order to specify this first-order relaxation, for each check $a \in
C$ we define a local codeword polytope (LCP): it is given by the
convex hull of the bit sequences that satisfy the check, which are
known as local codewords.  One way to define this LCP is as follows:
for each check $a \in C$, consider the set of bit sequences $z \in
\{0,1\}^\bitnum$ that \emph{fail} to satisfy check $a$, meaning that
$\oplus_{i \in a} z_i = 1$.  We denote this set of \emph{forbidden
sequences} by $\Forbid(a)$.  The local codeword polytope
$\operatorname{LCP}(a)$ associated with check $a$ consists of all
vectors $f \in [0,1]^\bitnum$ that are at Hamming distance at least
one from each $z \in \Forbid(a)$---viz.
\begin{eqnarray}
\label{EqnDefnLCP}
\operatorname{LCP}(a) & \defn & \left \{ f \in [0,1]^\bitnum \; \mid
\; \sum_{i \in N(a) \backslash \{k, z_k =1\} } f_i + \sum _{i\in \{k,
\; z_k = 1 \}} (1-f_i) \geq 1 \quad \forall \; z \in \Forbid(a) \right
\}.
\end{eqnarray}
(Note that any element $f_j$ with $j$ not in the neighborhood $N(a)$
of check $a$ is completely unconstrained.)  We refer to set of
$|Z(a)|$ inequality constraints defining the LCP as the
\emph{forbidden set inequalities}, and we refer to the $2 \bitnum$
inequalities $0\leq f_i \leq 1$ as the \emph{box inequality
constraints}.  Overall, the relaxed polytope $\mathcal{P}$ is defined
as the intersection of the LCPs over all checks---namely, $\mathcal{P}
\defn \cap_{a \in C} \operatorname{LCP}(a)$.  Note that for any check
$a$ with degree $d_c$, the number of local forbidden sequences is
$2^{d_c-1}$, so that for a check-regular code the total number of
forbidden sequences is $2^{d_c -1} \checknum$.  For low-density
parity-check codes, the maximum degree is bounded so that the relaxed
polytope can be described by a linear number of inequalities.  (For
higher degree checks, there is an alternative characterization of the
LCPs that is more efficient than the naive one described here; see
Feldman et al.~\cite{Feldman05} for details.)

If the LDPC graph has no cycles, the local forbidden sequences would
identify all the possible non-codewords, and the relaxation is
guranteed to be exact by a special case of the junction tree
theorem~\cite{WaiJor03Monster,WaiJaaWil05b}.  However, for a general
factor graph with cycles, there exist vertices with non $\{0,1\}$
coordinates that satisfy all the local constraints individually, and
yet are not codewords (nor linear combinations of codewords).  Such
sequences are called (fractional) pseudocodewords. To simplify the
presentation, we call the vertices of the relaxed polytope
pseudocodewords (so that codewords are also pseudocodewords).  The
term fractional pseudocodewords then designates the vertices of the
relaxed polytope that happen to have at least one fractional
coordinate.

\section{Structure of the relaxed polytope}
\label{SecStructural}

In this section, we turn to some theory concerning the structure of
the relaxed polytope.  In particular, we begin by addressing the
question of the minimal number of fractional coordinates in any
fractional pseudocodeword, a quantity that we term the fractional
support.  Although it is possible to construct codes with an
arbitrarily small fractional support, we show that for expander codes,
the fractional support has size linear in blocklength.  We then
address a second structural property of the polytope: namely, the
number of constraints that are active at any vertex.  By
dimensionality arguments, the size of this active set scales as
$\Theta(n)$.  Moving beyond this basic observation, we establish that
there is a dramatic difference between the active sets associated with
fractional pseudocodewords and those associated with (integral)
codewords.  More specifically, for expander codes, the active set of
any fractional pseudocodeword is smaller than the active set size of
any codeword by at least a constant fraction (in blocklength) of
constraints.  We leverage these structural results in
Section~\ref{SecImproved} to develop a randomized algorithm for
improving the performance of the LP-decoder by guessing facets of the
relaxed polytope and resolving the optimization problem.

\subsection{Fractional support of pseudocodewords}

The result of this section is to show that the fractional support of
any pseudocodeword in any LDPC code defined by an expander graph
scales linearly in blocklength.  We begin by defining the notion of an
\emph{expander graph}:
\begin{definition}  Given parameters $\alpha, \delta \in (0,1)$, we say that a
$(d_c, d_v)$-regular bipartite graph is an $(\alpha, \delta)$ expander
if, for all subsets $|S| \leq \alpha \nbit$, there holds $|\Neigh(S)|
\geq \delta d_v |S|$.
\end{definition}
Expander graphs have been studied extensively in past work on
coding~\cite{SipSpi96,bargzemor,Feldman07,DasDimKarWai06}.  Indeed, it is well-known
that randomly constructed regular LDPC graphs are expanders with high
probability (see, e.g.,~\cite{Feldman07}).

The fractional support of a pseudocodeword is
defined as follows.
\begin{definition}
The fractional support of a pseudocodeword $\pcw$ is the subset
$\vertfrac(\pcw) \subseteq \vertex$ of bits indices in which $\pcw$
has fractional coordinates.  Similarly, the subset of checks that are
adjacent to bits with fractional coordinates of $\pcw$ is denoted by
$\checkfrac(\pcw)$.
\end{definition}

The following result dictates that all fractional pseudocodewords in
an expander code have substantial fractional supports:
\begin{proposition}
\label{PropFracSupport}
Given an $(\alpha, \delta)$-expander code with \mbox{$\delta >
\frac{1}{2}$,} any pseudocodeword has fractional support that grows
linearly in blocklength:
\begin{eqnarray*}
|\vertfrac(\pcw)| \; \geq \; \alpha \nbit, \quad \mbox{and} \quad
|\checkfrac(\pcw)| \; \geq \; \delta \vdeg \alpha \nbit.
\end{eqnarray*}
\end{proposition}
\mybeginproof The proof exploits the following well-known
property~\cite{SipSpi96} of expander graphs.

\noindent \emph{Unique neighbor property:} Given an $(\alpha, \delta)$
expander with $\delta > \frac{1}{2}$, any subset $S \subseteq V$ of
size at most $\alpha n$ satisfies the unique neighbor property, i.e
there exists $y \in C$ such that $|\Neigh(y) \cap S|=1$.  To establish
this claim, we proceed via proof by contradiction: suppose that every
$y \in \Neigh(S)$ has two or more neighbors in $S$.  Then the total
number of edges arriving at $\Neigh(S)$ from $S$ is at least
\[
2|\Neigh(S)| > 2 \delta \vdeg |S| > \vdeg |S|.
\]
 But the total number of
edges leaving $S$ has to be exactly $\vdeg |S|$, which yields a
contradiction.

We now prove the stated proposition.  Consider any set $S$ of
fractional bits of size $|S| \leq \alpha \nbit$.  Using the expansion
and the unique neighbor property, the set $\Neigh(S)$ must contain at
least one check $a$ adjacent to only one bit in $S$.  However, we
claim that in any pseudocodeword $\pcw$, no check is adjacent to only
one fractional variable node.  Indeed, suppose that there were to
exist a check adjacent to only one fractional bit: then the associated
local pseudocodeword is in the local codeword polytope (LCP) for this
check and therefore can be written as a linear combination of two or
more codewords~\cite{Ziegler}.  But these local codewords would have
to differ in only one bit, which is not possible for a parity check.

Therefore, the check $a$ must be adjacent to at least one additional
fractional bit (not in $S$).  We then add this bit to $S$, and repeat
the above argument until $|S| > \alpha \nbit$, to conclude that
$|\vertfrac(\pcw)| > \alpha \nbit$.  Finally, the bound on
$|\checkfrac(\pcw)|$ follows by applying the expansion property to a
subset of fractional bits of size less than or equal to $\alpha
\nbit$.
\myendproof

{\bf{Remark:}} In fact, a careful examination of the proof reveals
that we can make a slightly stronger claim.  Given a pseudocodeword
with fractional support $S$, consider the graph $\graph[S]$ induced by
the fractional bits, which may have multiple connected components.
The proof of Proposition~\ref{PropFracSupport} shows that the size of
every connected component must grow linearly in the blocklength for
suitable expander codes.

\subsection{Sizes of active sets}

For any vertex $v$ of a polytope, its active set $\actset(v)$ is the
set of linear inequalities that are satisfied with equality on
$v$. Geometrically, this corresponds to the set of facets of the
polytope that contain the vertex $v$.  For LP decoding, the set of
possible vertices includes both codewords and (fractional)
pseudocodewords.  The key property that we prove in this section is
that for expander codes, codewords have active sets which are larger
by at least a constant factor than the active sets of fractional
pseudocodewords.

\begin{figure}
\begin{center}
\begin{tabular}{cc}
\psfrag{#ac#}{$\actset(\cw)$} \psfrag{#ap#}{$\actset(\pcw)$}
\psfrag{#vf#}{$\vertex(F)$} \psfrag{#vert#}{vertices}
\psfrag{#fac#}{facets}
\widgraph{.28\textwidth}{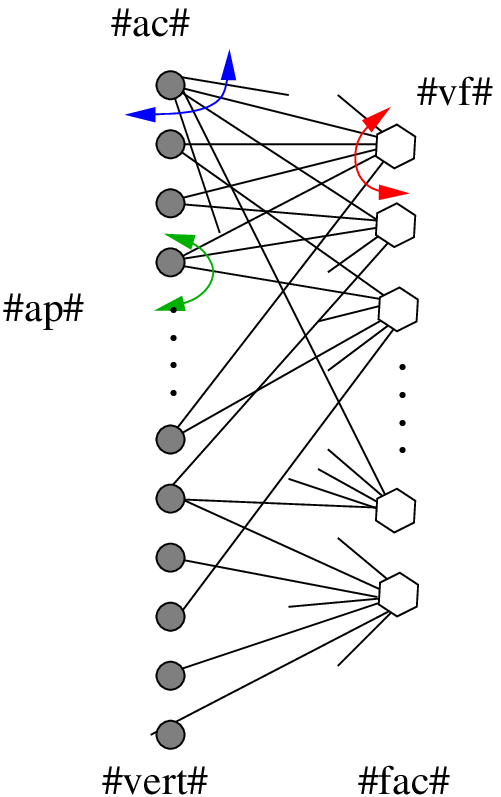} &
\psfrag{#hl#}{$\mbox{Higher likelihood}$}
\psfrag{#aml#}{$\actset(\xml)$}
\psfrag{#xml#}{$\xml$}
\psfrag{#F*#}{$F^*$}
\widgraph{.28\textwidth}{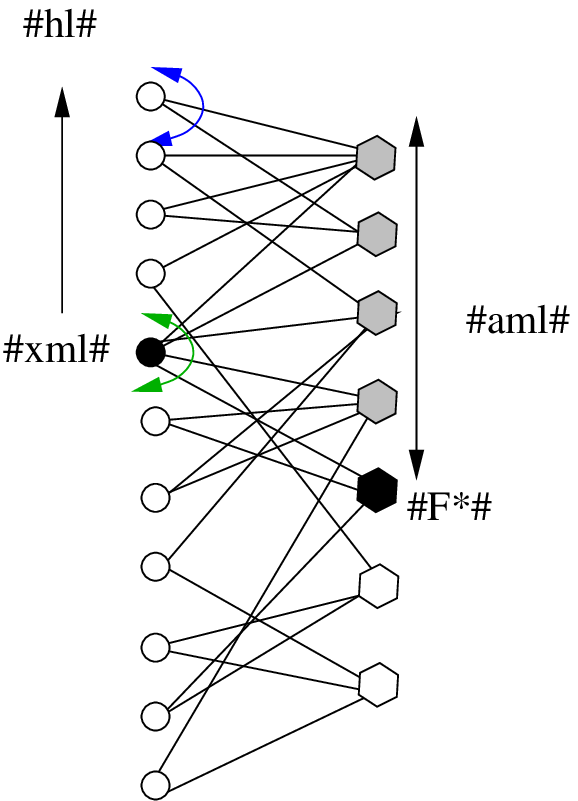} \\
(a) & (b)
\end{tabular}
\end{center}
\caption{(a) Vertex-facet diagram of the relaxed polytope.  Circles on
the left-hand side (LHS) correspond to vertices (codewords $\cw$ and
fractional pseudocodewords $\pcw$) of the relaxed polytope; hexagons
on the right-hand side (RHS) correspond to facets (hyperplane
inequalities) defining the relaxed polytope. The neighborhood of the
LHS node associated with a pseudocodeword $\pcw$ defines its active
set $\actset(\pcw)$.  (b) Illustration of proof of
Corollary~\ref{CorExhaust}.  Vertices (codewords and pseudocodewords)
are ordered by likelihood on LHS. Facet-guessing algorithm fails only
if the active set $\actset(\xml)$ of the ML codeword is fully covered
by the active sets $\actset(\pcw)$ of pseudocodewords with higher
likelihood.  Here facet $F^*$ remains uncovered so that the algorithm
succeeds.}
\label{FigVertexFacet}
\end{figure}

Before stating and proving this result, let us introduce the
vertex-facet diagram~\cite{Ziegler} that describes the relation
between the polytope vertices and facets.  This diagram can be
understood as a bipartite graph $B$ with the set of all codewords and
pseudocodewords (vertices of $\mathcal{P}$) on its left-hand side, and
the set of all constraints (facets of $\mathcal{P}$) on its right-hand
side.  Any given (pseudo)codeword $\pcw$ is connected to a given facet
$F$ if and only if $\pcw \in F$; see Figure~\ref{FigVertexFacet} for
an illustration.  In this diagram, the active set $\actset(\pcw)$ of a
given pseudocodeword $\pcw$ is simply the set of neighbors of the LHS
node representing $\pcw$.  The main result of this section concerns
the degrees of the LHS nodes, or the sizes $|\actset(\pcw)|$ and
$|\actset(\cw)$ of the (fractional) pseudocodeword and codeword active
sets.

\begin{atheorem}
\label{ThmPCWSize} For any $(\vdeg, \cdeg)$ code with $\rate \in
(0,1)$, the active set of any codeword $\cw$ has
\begin{eqnarray}
\label{EqnActsetCW} |\actset(\cw)| & = \gamma_{cw} \nbit.
\end{eqnarray}
elements.  For an $(\alpha, \delta)$-expander code with $\delta >
\frac{1}{2}$, the active set of any fractional pseudocodeword $\pcw$
is smaller than the active set of any codeword by a linear
fraction---in particular,
\begin{eqnarray}
\label{EqnActsetPCW} |\actset(\pcw)| & \leq & \nbit \gamma_{pc}
 % \nonumber \\
%%
%& = & |\actset(\cw)| - \nbit \biggr \{ \alpha \big[ \vdeg \delta
%  \big(\cdeg-2 \big) +1 \big] \biggr\}.
\end{eqnarray}
where the constants are \mbox{$\gamma_{cw}=\big[ (1-\rate )\cdeg +1
\big]$} and \mbox{$\gamma_{pc}= \Big[ \big( 1- \rate -\delta \vdeg
\alpha \big) \cdeg + 2 \delta \vdeg \alpha + (1 - \alpha)\Big]$.}
(Note that $\gamma_{pc} < \gamma_{cw}$.)
\end{atheorem}
\mybeginproof
We begin by proving equation~\eqref{EqnActsetCW}.  By the
code-symmetry of the relaxed polytope~\cite{Feldman05}, every codeword
has the same number of active inequalities, so it suffices to restrict
our attention to the all-zeroes codeword.  The check inequalities
active at the all-zeros codeword are in one-to-one correspondence with
those forbidden sequences at Hamming distance $1$.  Note that there
are $\cdeg$ such forbidden sequences, so that the total number of
constraints active at the all-zeroes codeword is simply
\begin{equation*}
|\actset(\cw)| = \mcheck \cdeg + \nbit \; = \; \nbit \, \big[
(1-\rate) \cdeg + 1 \big],
\end{equation*}
as claimed.

We now turn to the proof of the bound~\eqref{EqnActsetPCW} on the size
of the fractional pseudocodeword active set.  Recall that the relaxed
polytope consists of two types of inequalities: \emph{forbidden set
constraints} (denoted $\Forbid$) associated with the checks, and the
\emph{box inequality constraints} $0 \leq x_i \leq 1$ (denoted
$\Boxin$) associated with the bits.  The first ingredient in our
argument is the fact (see Proposition~\ref{PropFracSupport}) that for
an $(\alpha, \delta)$-expander, the fractional support
$\vertfrac(\pcw)$ is large, so that a constant fraction of the box
inequalities will not be active.

Our second requirement is a bound on the number of forbidden set
inequalities that can be active at a pseudocodeword.  We establish a
rough bound for this quantity using the following lemma:
\begin{lemma}
\label{LemSurf1}
Suppose that $z$ belongs to a polytope and is not a vertex.  Then
there always exist at least two vertices $x, y$ such that $\actset(z)
\subseteq \actset(x) \cap \actset(y)$.
\end{lemma}
\mybeginproof Since $z$ belongs to the polytope but is not a vertex,
it must either belong to the interior, or lie on a face with dimension
at least one.  If it lies in the interior, then \mbox{$\actset(z)=
\emptyset$,} and the claim follows immediately.  Otherwise, $z$
must belong to a face $F$ with $\dim(F) \geq 1$.  Then $F$ must
contain~\cite{Ziegler} at least $\dim(F) + 1 = 2$ vertices, say $x$
and $y$.  Consequently, since $x, y$ and $z$ all belong to $F$ and $z$
is not a vertex, we must have $\actset(z) \subseteq \actset(y)$ and
$\actset(z) \subseteq \actset(x)$, which yields the claim.
\myendproof

Given a check $c$ and codeword $\cw$, let $\Pi_c(\cw)$ denote the
restriction of $\cw$ to bits in the neighborhood of $c$ (i.e., a
\emph{local codeword} for the check $c$).  With this notation, we
have:
\begin{lemma}
\label{LemSurf2}
For any two local codewords \[ \Pi_c(\cw_1) , \Pi_c(\cw_2)\] of a
 check $c$, the following inequality holds
\[
|\actset(\Pi_c(\cw_1)) \cap \actset(\Pi_c(\cw_2))| \leq 2.
\]
\end{lemma}
\mybeginproof
The intersection
\[
\actset(\Pi_c(\cw_1)) \cap \actset(\Pi_c(\cw_2))
\]
is
given by the forbidden sequences that have Hamming distance $1$ from
$\Pi_c(\cw_i), i = 1,2$ (i.e., forbidden sequences $f$ such that $d(f,
\Pi_c(\cw_i)) = 1$ for $i=1,2$).  Thus, if such an $f$ exists, then by
the triangle inequality for Hamming distance, we have
\[
2 = d(f,\Pi_c(\cw_1)) + d(f, \Pi_c(\cw_2))) \geq d(\Pi_c(\cw_1), \Pi_c(\cw_2)),
\]
But 
\[
d(\Pi_c(\cw_1), \Pi_c(\cw_2)) \geq 2\]
for any two local
codewords, so that we must have 
\[
d(\Pi_c(\cw_1), \Pi_c(\cw_2)) = 2.
\]
Consequently, we are looking for all the forbidden (odd) sequences of
length $\cdeg$ that differ in one bit from two local codewords that
are different in two places.  Clearly there are only two such
forbidden sequences, so that the claim follows.
\myendproof \\

We can now establish a bound on the size of the active sets of
pseudocodewords for $(\alpha, \delta)$-expanders:
\begin{lemma}
\label{LemUpperBound}
For every pseudocodeword $\pcw$, the size of the active set
$|\actset(\pcw)|$ is upper
bounded by
\begin{equation}
\label{EqnActiveSet}
(\mcheck -|\checkfrac(\pcw)|) \cdeg + 2
|\checkfrac(\pcw)|
+ \nbit - |\vertfrac(\pcw)|.
\end{equation}
\end{lemma}
\mybeginproof The proof is based on the decomposition:
\begin{eqnarray*}
|\actset(\pcw)| & = & |\actset(\pcw) \cap \Forbid| + | \actset(\pcw)
\cap \Boxin|.
\end{eqnarray*}
The cardinality
$|\actset(\pcw) \cap \Boxin|$ is equal to the number
of integral bits in the pseudocodeword, given by $\nbit -
|\vertfrac(\pcw)|$.  We now turn to upper bounding the cardinality
$|\actset(\pcw) \cap \Forbid|$.  Consider the $\mcheck -
|\checkfrac(\pcw)|$ checks that are adjacent to only integral bits of
$\pcw$.  For each such check, exactly $\cdeg$ forbidden set
constraints are active, thereby contributing a total of
\[
\cdeg \big[\mcheck - |\checkfrac(\pcw)|\big]
\]
 active constraints.  Now
consider one of the remaining $|\checkfrac(\pcw)|$ fractional checks,
say $c$.  Consider the restriction $\Pi_c(\pcw)$ of the pseudocodeword
$\pcw$ to the check neighborhood of $c$.  Since $\Pi_c(\pcw)$ contains
fractional elements, it is not a vertex of the local codeword polytope
associated with $c$.  Therefore, by combining Lemmas~\ref{LemSurf1}
and~\ref{LemSurf2}, we conclude that
\[
|\actset(\Pi_c(\pcw))| \leq 2.
\]
Overall, we conclude that the upper bound~\eqref{EqnActiveSet} holds.
\myendproof \\

Using Lemma~\ref{LemUpperBound} and Proposition~\ref{PropFracSupport},
we can now complete the proof of Theorem~\ref{ThmPCWSize}.  In
particular, we re-write the RHS of the bound~\eqref{EqnActiveSet} as
%\begin{equation*}
%\mbox{$(1-\rate)\cdeg \, \nbit - (\cdeg - 2) |\checkfrac(\pcw)| +
%  \nbit - |\vertfrac(\pcw)|$.}
%\end{equation*}

\begin{equation*}
(1-\rate)\cdeg \, \nbit - (\cdeg - 2) |\checkfrac(\pcw)| +
  \nbit - |\vertfrac(\pcw)|.
\end{equation*}

From Proposition~\ref{PropFracSupport}, we have $|\checkfrac(\pcw)|
\geq \vdeg \delta \alpha \nbit$ and $|\vertfrac(\pcw)| > \alpha
\nbit$, from which the bound~\eqref{EqnActsetPCW} follows.
\myendproof

\section{Improved LP decoding}
\label{SecImproved}

Various improved decoding algorithms have been suggested in past work,
both based on extensions of standard iterative
decoding~\cite[e.g.,]{Fossorier01,PishroFekri} as well as extensions
of LP decoding~\cite{CheChe06,YanFelWan06}.  Based on the structural
results that we have obtained, we now describe an improved decoding
algorithm for which some finite-length theoretical guarantees can be
made.  We begin with some simple observations: (i) ML decoding
corresponds to finding the vertex in the relaxed polytope that has the
highest likelihood and integral coordinates; and (ii) Standard LP
decoding succeeds if and only if the ML codeword has the highest
likelihood over all pseudocodewords.

These observations highlight the distinction between LP decoding and
ML decoding.  An LP solver, given the (polynomially many) facets of
the relaxed polytope, determines the vertex with the highest
likelihood without having to go through all the exponentially many
vertices of $V$. In contrast, the ML decoder can go down this list,
and determine the first vertex which has integral coordinates.  This
motivates facet-guessing: suppose that there exists only one
fractional pseudocodeword $\pcw_1$ that has higher likelihood than the
ML codeword $\xml$.  The LP decoder will output the pseudocodeword
$\pcw_1$, resulting in a decoding error.  However, now suppose that
there exists a facet $F_1 \in \actset$ such that $\xml \in F_1$ but
$\pcw \notin F_1$.  Consider the reduced polytope $\mathcal{P'}$
created by restricting the relaxed polytope $\mathcal{P}$ to the facet
$F_1$ (i.e., $\mathcal{P'} = \RelPoly \cap F_1$).  This new polytope
will have a vertex-facet graph $\mathcal{B'}$ with vertices $V'=
N(F_1)$ i.e. all the vertices that are contained in $F_1$. The
likelihoods will be the same, but $p_1$ will not belong in
$\mathcal{P'}$ and therefore we can use an LP solver to determine the
vertex with the highest likelihood in $\mathcal{P'}$.  If we had
chosen the correct facet, this vertex would be the ML codeword $\xml$.
Based on this intuition, we now formally describe the facet-guessing
algorithm for improved LP decoding. \\

\framebox[0.98\textwidth]{\parbox{.95\textwidth}{
{\bf{Facet Guessing Algorithm}}
\begin{enumerate}
\item Run LP decoding: if outputs an integral codeword, terminate.
Otherwise go to Step 2.
\item Take as input:
\begin{itemize}
\item fractional pseudocodeword $\pcw$ from the LP decoder
\item likelihood vector $\gamma$.
\end{itemize}

\item Given a natural number $N \geq 1$, repeat for $i=1, \ldots N$:
\begin{enumerate}
\item[(a)] Select a facet $F_i \in (\actset \setminus \actset(\pcw)$,
  form the reduced polytope $\mathcal{P'} = \RelPoly \cap F_i$.

\item[(b)] Solve the linear program with objective vector $\gamma$ in
$\mathcal{P'}$, and save the optimal vertex $z_i$.
\end{enumerate}

\item From the list of optimal LP solutions $\{z_1, \ldots, z_N \}$,
output the integral codeword with highest likelihood.
\end{enumerate}
}} \\
%%%%%%%%%%%%%%%%%%%%%%%%%%%%

\myparagraph{Remarks:}

\begin{enumerate}
\item[(a)] We can consider two variations of facet guessing:
exhaustive facet guessing (EFG) tries all possible facets (i.e., $N =
|(\actset \setminus \actset(\pcw))|$), while randomized facet guessing
(RFG) randomly samples from $(\actset \setminus \actset(\pcw))$ a
constant number of times (e.g., $N = 20$).
\item[(b)] Regardless of the problems, the exhaustive facet-guessing
(EFG) algorithm has polynomial-time complexity, since the number of
calls to the LP solver grows linearly as 
\[
|\actset \setminus \actset(\pcw)| = \mathcal{O}(n).
\]
On the other
hand, the RFG algorithm requires only a constant number of calls to an
LP solver, and therefore has the same order
of polynomial complexity as standard LP decoding.  When these
algorithms are applied to a sequence of decoding problems, one would
expect that the average complexity is typically very close to LP
decoding, since the facet-guessing routines (Step 2) run \emph{only
if} the LP decoder has already failed.
\end{enumerate}

We now provide a simple necessary and sufficient characterization for
the EFG algorithm to fail:
\begin{lemma}
The exhaustive facet-guessing algorithm fails to find the ML codeword
$\xml$ $\iff$ every facet $F \in \actset(\xml)$ contains a fractional
pseudocodeword with likelihood greater than $\xml$.
\label{exhaustive}
\end{lemma}
\mybeginproof Denote the set of fractional pseudocodewords with
likelihood higher than $\xml$ by $\mathbb{P}(\xml)$. Assume there
exists a facet $F_i$ such that $\xml \in F_i$ and $\pcw \notin F_i$
for all pseudocodewords $\pcw \in \mathbb{P}(\xml)$.  Then the
facet-guessing algorithm will at some round select the facet $F_i$,
and the LP solver will output $\xml$, as the vertex in $\mathcal{P'}$
with the highest likelihood.  Consequently, the ML solution $\xml$
will be in the list of LP solutions in step (4).  Since $\xml$ is the
ML codeword, there can be no other integral codeword with higher
likelihood in the list, so that the algorithm must output $\xml$.
Conversely, suppose that every facet $F \in \actset(\xml)$ contains a
fractional pseudocodeword with likelihood greater than $\xml$.  Then,
the ML codeword $\xml$ will never be the output of the LP solver at
any round, since some pseudocodeword will always have higher
likelihood.  Consequently, the ML codeword will not appear in the
final list, so that the facet-guessing method must fail.  \myendproof

\begin{figure*}
\begin{center}
\begin{tabular}{cc}
\widgraph{.48\textwidth}{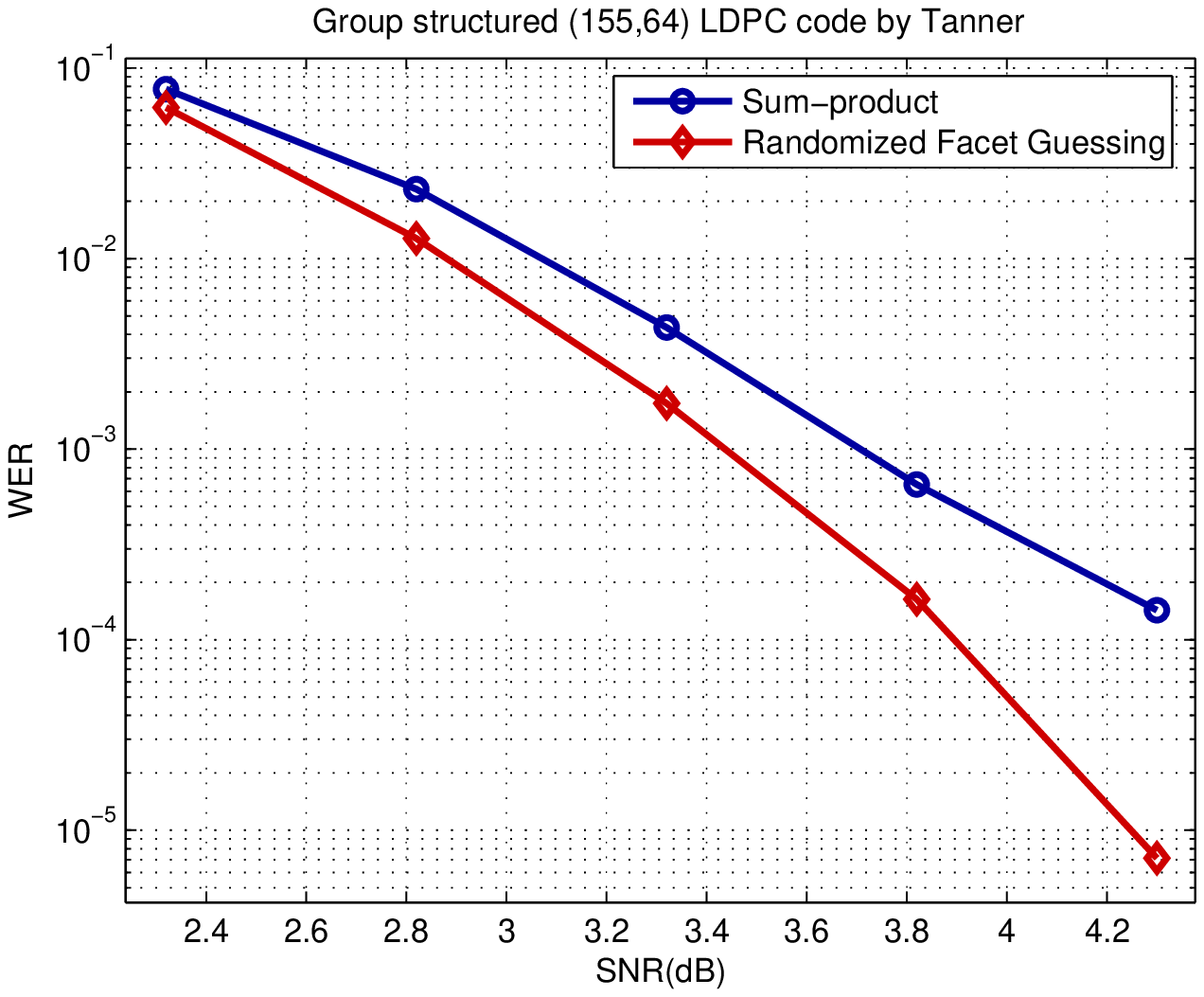} &
\widgraph{.48\textwidth}{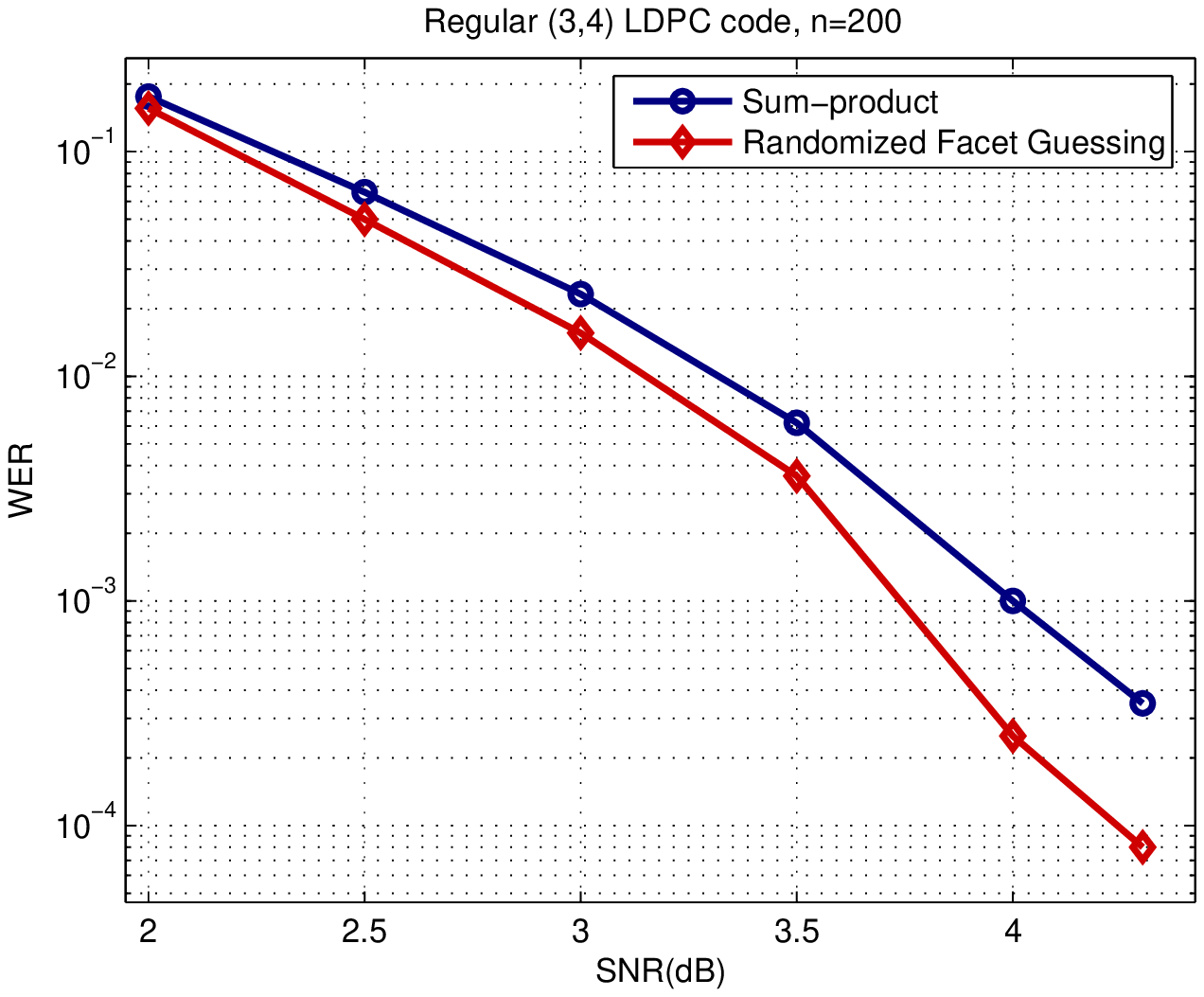} \\ (a) & (b)
\end{tabular}
\end{center}
\caption{Comparison of different decoding methods: standard
sum-product decoding, and randomized facet-guessing (RFG) with $N =
20$ iterations. The two panels show two different codes: (a) Tanner's
group-structured code.  (b) Random (3,4) LDPC code with $n=200$. }
\label{FigPerformance}
\end{figure*}

We now combine this characterization of the success/failure of
facet-guessing with our earlier structural properties of expander
codes.  Doing so yields the following result:
\begin{corollary}
\label{CorExhaust}
For expander codes, the EFG algorithm will
always succeed if there are $\const_1 \frac{\gamma_{cw}}{\gamma_{pc}}$
fractional pseudocodewords with likelihood higher than the ML
codeword.  Moreover, under this same condition, each iteration of RFG
succeeds with constant probability
\[
p_{RFG} \geq \frac{
\gamma_{cw} - \const_1 \gamma_{pc}} {2^{d_c-1}(1-R)+2}.
\]
%
%will always succeed if there are less than or equal to
%\begin{eqnarray}
%\const_1 & \defn & \left \lfloor \frac{(1-\rate)d_c +1}{ \big(
%1-\rate - \alpha d_v/2 \big)d_c +d_v \alpha + (1 - \alpha)} \right
%\rfloor
%\end{eqnarray}
%fractional pseudocodewords with higher likelihood.
\end{corollary}
\mybeginproof
From Lemma~\ref{exhaustive}, the EFG algorithm fails if and only if
every facet in $|\actset_c|$ also contains another fractional
pseudocodeword with higher likelihood.  But for expander codes,
Lemma~\ref{LemUpperBound} yields that the size of the active set of
any fractional pseudocodeword is upper bounded as
\begin{eqnarray*}
|\actset_p| & \leq n \gamma_{pc}.
\end{eqnarray*}
while the size of active sets of any codeword is always $|\actset_c|=
n \gamma_{cw}$. Therefore, if there exist $\const_1$ fractional
pseudocodewords with likelihood higher than $c$, the total number of
facets adjacent to these fractional pseudocodewords is at most
\[
\gamma_{pc} \const_1 n.
\]
Therefore when
\[
\gamma_{pc} \const_1 n < n \gamma_{cw},
\]
it is impossible to completely cover $\actset_c$ and EFG
succeeds. Also RFG at each iteration selects a random facet and there
are
\[
(\gamma_{cw} -\gamma_{pc}\const_1) n
\]
facets that contain $c$ but
not any fractional pseudocodeword with higher likelihood. The total
number of facets is
\[
|\actset|= (2^{d_c-1}(1-R)+2) n
\] and therefore
each iteration of RFG has probability of success larger than
\[
\frac{ \gamma_{cw} - \const_1 \gamma_{pc}} {2^{d_c-1}(1-R)+2}.
\]  \myendproof
\\

Notice that this corollary only provides a worst case bound.  Indeed,
the bound is achieved in a somewhat unlikely manner: it requires a set
of fractional pseudocodewords all with higher likelihood than the ML
codeword $\xml$, all of whose active sets are entirely contained
within the active set $\actset(\xml)$ of the ML codeword, and all of
whose active sets are pairwise disjoint.  (See
Figure~\ref{FigVertexFacet}(b) for an illustration.)  More typically,
one could expect the facet guessing algorithm to work even if there
are many more fractional pseudocodewords with higher likelihoods.
Indeed, our experimental results show that the RFG algorithm leads to
a significant performance gain for those codewords, frequently
correctly recovering the ML codeword in cases for which both
sum-product and LP decoding fail.  As shown in
Figure~\ref{FigPerformance}, the gains are pronounced for higher SNR,
as high as $0.5$dB for the small blocklengths that we experimentally
tested.  The added complexity corresponds to solving a constant number
of LP optimizations; moreover, the extra complexity is required
\emph{only if} LP decoding fails.

%%%%%%%%%%%%%%%%%%%%%%%%%%%%%%%%%%%%%%%%%%%%%%%%%%%%%%%%%%%%%%

%--------------------------------
\section{Improved theoretical guarantees}
\label{SecHigher}

The facet-guessing algorithm described in the previous section is
based on choosing facets at random. Note that the chosen facet may
either be of the \emph{forbidden set type}, or the \emph{box
constraint type} (e.g., $0 \leq f_i \leq 1$).  In this section, we
describe a particular type of facet-guessing algorithm that chooses only box
inequalities, and hence has the natural interpretation of a
bit-guessing algorithm~\cite{PishroFekri}.  We
show how this facet-guessing algorithm can be used to find the ML codeword
in polynomial time, as long as the number
of fractional pseudocodewords with higher likelihood is bounded by a polynomial.

\subsection{Random bit guessing}
The basic intuition underlying our random bit-guessing algorithm is
simple.  Since the LP decoding algorithm runs in polynomial time, we can afford to solve up to a polynomial number of
linear programs to decode.
 Accordingly, we propose to
choose a subset of bits, with size scaling as $c \log{n}$ for some
constant $c>0$, and to try all possible $2^{c\log{n}}=n^c$ possible
$0-1$ configurations indexed by bits in this subset.  In one of these
trials, the chosen configuration of $c\log{n}$ bits will match with
the corresponding bits in the ML codeword.  The algorithm will only
fail if a ``bad'' pseudocodeword happens to coincide with the ML
codeword in all $c\log{n}$ positions.  The formal description of the
algorithm is as follows: \\

\framebox[0.98\textwidth]{\parbox{.95\textwidth}{ {\bf{Randomized bit
guessing (RBG) algorithm}}
\begin{enumerate}
\item Run LP decoding: terminate if it outputs an integral codeword;
otherwise go to step 2.
\item Choose $c\log{n}$ bits (randomly or deterministically),
$x_{i_1}$, $x_{i_2}$, $x_{i_3}$, \ldots, $x_{i_{c\log{n}}}$.

\item Take as input the likelihood vector $\gamma$, and repeat the
following loop for $2^{c\log{n}} = n^c$ times:
\begin{enumerate}
\item[(a)] Consider a new 0-1 configuration out of the total
$2^{c\log{n}}$ configurations for the $c\log{n}$ bits, say
($\alpha_{i_1}$, $\alpha_{i_2}$, $\alpha_{i_3}$, \ldots,
$\alpha_{i_{c\log{n}}}$).

\item[(b)] Add the equations $x_{i_j}=\alpha_{i_j}$ for $j=1 \ldots
c\log{n}$ to the set of inequalities defining the relaxed polytope and
solve the linear program with objective vector $\gamma$ in
$\mathcal{P'}$. Save the optimal vertex $z_i$.
\end{enumerate}

\item From the list of optimal LP solutions $\{z_1, \ldots, z_{c\log{n}} \}$,
output the integral codeword with highest likelihood.
\end{enumerate}
}}

\subsection{Analysis}

Suppose that the set of $c\log{n}$ bits are chosen randomly. The main
theorem of this section shows that this random bit-guessing (RBG)
algorithm succeeds if there any at most polynomially many ``bad''
pseudocodewords.  More formally, we let $M$ denote the number of
pseudocodewords $\pcw$ that
\begin{enumerate}
\item[(a)] have higher likelihood than the ML codeword $\xml$, and
\item[(b)] are adjacent to ML-codeword on the relaxed codeword
polytope, meaning that the intersection $\actset(\xml) \cap
\actset(\pcw)$ is non-empty.
\end{enumerate}
With this definition, we have the following:
\begin{atheorem}
\label{ThmPCWSize} Given an $(\alpha, \delta)$-expander code with
\mbox{$\delta > \frac{1}{2}$}, the RBG algorithm finds the ML-codeword
with probability \[1- M/(n^{-c\log{(1-\alpha)}}).\]  Consequently, for
any order $M = \mathcal{O}(n^b)$ of polynomial growth, the RBG
algorithm succeeds with probability converging to one for all \[c
> b/\log(1-\alpha).\]
\end{atheorem}
\mybeginproof By the code symmetry of the relaxed
polytope~\cite{Feldman05}, we may assume without loss of generality
that $\xml$ is the all-zeroes codeword (although the algorithm does
not know this information).  If $\xml$ is the all-zeroes word, then
the key iteration of the RBG algorithm is the step at which it sets
$x_{i_j}=0$ for $j=1 \ldots c\log{n}$.
From
Proposition~\ref{PropFracSupport}, since the graph is an $(\alpha,
\delta)$-expander code, every pseudocodeword has at least $\alpha n$
fractional coordinates. Therefore, a randomly chosen bit from any
pseudocodeword will be integral with probability at most $1-\alpha$.
Consequently, if we force a set $c\log{n}$ bits to zero (as in the key
step described above), then the probability that all the bits fall
outside the fractional support of any given pseudocodeword is at most
\[(1-\alpha)^{c\log{n}}.\]  Otherwise stated, with probability at least
\[1-(1-\alpha)^{c\log{n}},\] a random selection of $c\log{n}$ bits will
exclude any particular pseudocodeword as a possible output of the RBG
algorithm.  By a union bound, any set of $M$ pseudocodewords are
excluded with probability at least
\[
1- M(1-\alpha)^{c\log{n}} = 1- M n^{c\log{(1-\alpha)}}.
\]
Consequently, if there at at most $M$
pseudocodewords with likelihood higher than the ML codeword, then the
RBG algorithm will succeed with at least this probability.

In order to complete the proof, we need to show that it is sufficient
to exclude only higher likelihood pseudocodewords that are also
adjacent on the relaxed polytope $\mathcal{P}$ to the all-zeroes ML
codeword.  In order for the all-zeroes $\xml$ to \emph{not} be the
output of the restricted LP at the key step (in which the set of $c
\log n$ bits are set to zero), there must exist a pseudocodeword in
the restricted polytope
\[
\mathcal{P} \cap \left(\cap_{j=1}^{c \log n} \{x_{i_j}=0\} \right)
\]
with higher likelihood.  Any such
pseudocodeword is certainly adjacent to the all-zeroes codeword, since
they share all the box constraints $x_{i_j} = 0$.  Therefore, it is
sufficient to exclude only ``bad'' pseudocodewords that are adjacent
to the ML-codeword on the relaxed polytope.  \myendproof

\section{Conclusions}
\label{SecDiscussion}

In this paper, we have investigated the structure of the polytope that
underlies both the LP method and the sum-product algorithm for
decoding of low-density parity check codes.  For codes based on
suitable expander graphs, we proved a number of structural properties
of this polytope, including the fact that any (fractional)
pseudocodeword has at least a constant fraction of non-integral bits,
and that the number of active sets differ substantially between
pseudocodewords and codewords.  Inspired by these structural
properties, we proposed a number of efficient decoding algorithms that
offer quantifiable improvements over basic LP decoding.  First, we
described a facet-guessing algorithm that has complexity equivalent
(apart from a constant factor) to standard LP decoding, and provided
both theoretical and empirical results on the performance gains that
it achieves.  We also proposed a randomized bit-guessing algorithm,
and proved that it can still recover the ML codeword as long as there
are at most a polynomial number of pseudocodewords with higher
likelihood.

The results of this paper raise an interesting question concerning the
structure of pseudocodewords in various code families.  Previous work
by Koetter and Vontobel~\cite{KoeVon03} established that for any
bit-check regular LDPC code, there exist pseudocodewords for the
additive white Gaussian noise (AWGN) channel with sublinear weight.
This fact implies that standard LP decoding cannot have an error
exponent for the AWGN, meaning an exponential decay in error
probability.\footnote{Although subsequent work~\cite{FelKoeVon05}
showed that LP decoding does have an error exponent if the log
likelihoods are suitably thresholded, but this procedure discards
potentially useful information.}  While standard LP decoding can be
compromised by a single ``bad'' pseudocodeword, the improved decoding
procedures in this paper are still guaranteed to recover the ML
codeword even if there are a polynomial number of pseudocodewords with
sublinear weight.  Therefore, it would be interesting to determine
which code families do (or do not) have a super-polynomial number of
sublinear weight pseudocodewords.


\section*{Acknowledgment}
Work partially supported by NSF Grant DMS-0528488, NSF Grant CAREER
CCF-0545862, and a UC-MICRO grant through Marvell Semiconductor.

% http://www.ctan.org/tex-archive/macros/latex/contrib/supported/IEEEtran/bibtex
%\bibliographystyle{IEEEtran.bst}

\bibliographystyle{IEEEtran}

\bibliography{BIBTEX_a_isit06}

\begin{thebibliography}{10}
\providecommand{\url}[1]{#1}
\csname url@rmstyle\endcsname
\providecommand{\newblock}{\relax}
\providecommand{\bibinfo}[2]{#2}
\providecommand\BIBentrySTDinterwordspacing{\spaceskip=0pt\relax}
\providecommand\BIBentryALTinterwordstretchfactor{4}
\providecommand\BIBentryALTinterwordspacing{\spaceskip=\fontdimen2\font plus
\BIBentryALTinterwordstretchfactor\fontdimen3\font minus
  \fontdimen4\font\relax}
\providecommand\BIBforeignlanguage[2]{{%
\expandafter\ifx\csname l@#1\endcsname\relax
\typeout{** WARNING: IEEEtran.bst: No hyphenation pattern has been}%
\typeout{** loaded for the language `#1'. Using the pattern for}%
\typeout{** the default language instead.}%
\else
\language=\csname l@#1\endcsname
\fi
#2}}

\bibitem{Gallager63}
R.~G. Gallager, \emph{Low-density parity check codes}.\hskip 1em plus 0.5em
  minus 0.4em\relax Cambridge, MA: MIT Press, 1963.

\bibitem{Richardson01a}
T.~Richardson and R.~Urbanke, ``The capacity of low-density parity check codes
  under message-passing decoding,'' \emph{IEEE Trans. Info. Theory}, vol.~47,
  pp. 599--618, February 2001.

\bibitem{Ashikhmin04}
A.~Ashihkmin, G.~Kramer, and S.~ten Brink, ``Extrinsic information transfer
  functions: model and erasure channel properties,'' \emph{IEEE Trans. Info.
  Theory}, vol.~50, no.~11, pp. 2657--2673, 2004.

\bibitem{Feldman05}
J.~Feldman, M.~J. Wainwright, and D.~R. Karger, ``Using linear programming to
  decode binary linear codes,'' \emph{IEEE Transactions on Information Theory},
  vol.~51, pp. 954--972, March 2005.

\bibitem{Feldman03}
J.~Feldman, D.~R. Karger, and M.~J. Wainwright, ``Using linear programming to
  decode {LDPC} codes,'' in \emph{Conference on Information Science and
  Systems}, March 2003.

\bibitem{FelKarWai02}
------, ``Linear programming-based decoding of turbo-like codes and its
  relation to iterative approaches,'' in \emph{Proc. 40th Annual Allerton Conf.
  on Communication, Control, and Computing}, October 2002.

\bibitem{Wainwright02aller}
M.~J. Wainwright, T.~S. Jaakkola, and A.~S. Willsky, ``{MAP} estimation via
  agreement on (hyper)trees: Message-passing and linear programming
  approaches,'' in \emph{Proc. Allerton Conference on Communication, Control
  and Computing}, October 2002.

\bibitem{WaiJaaWil05b}
------, ``Exact {MAP} estimates via agreement on (hyper)trees: Linear
  programming and message-passing,'' \emph{IEEE Trans. Information Theory},
  vol.~51, no.~11, pp. 3697--3717, November 2005.

\bibitem{KoeVon03}
R.~Koetter and P.~O. Vontobel, ``Graph-covers and iterative decoding of finite
  length codes,'' in \emph{Proc. 3rd International Symp. on Turbo Codes},
  September 2003.

\bibitem{DasDimKarWai06}
C.~Daskalakis, A.~G. Dimakis, R.~M. Karp, and M.~J. Wainwright, ``Probabilistic
  analysis of linear programming decoding,'' in \emph{Proceedings of the 18th
  Annual Symposium on Discrete Algorithms (SODA)}, January 2007.

\bibitem{Feldman07}
J.~Feldman, T.~Malkin, R.~A. Servedio, C.~Stein, and M.~J. Wainwright, ``{LP}
  decoding corrects a constant fraction of errors,'' \emph{IEEE Trans.
  Information Theory}, vol.~53, no.~1, pp. 82--89, January 2007.

\bibitem{kv_bethe}
P.~Vontobel and R.~Koetter, ``Lower bounds on the minimum pseudo-weight of
  linear codes,'' in \emph{International Symposium on Information Theory (ISIT
  '04), Chicago, IL}, June 2004.

\bibitem{FelKoeVon05}
J.~Feldman, R.~Koetter, and P.~O. Vontobel, ``The benefit of thresholding in
  {LP} decoding of {LDPC} codes,'' in \emph{International Symposium on
  Information Theory}, 2005, pp. 307--311.

\bibitem{VonKoe06b}
P.~O. Vontobel and R.~Koetter, ``Towards low-complexity linear-programming
  decoding,'' in \emph{Proc. Int. Conf. on Turbo Codes and Related Topics},
  Munich, Germany, April 2006.

\bibitem{TagSie06}
M.~H. Taghavi and P.~H. Siegel, ``Adaptive linear programming decoding,'' in
  \emph{IEEE Int. Symposium on Information Theory}, Seattle, WA, July 2006.

\bibitem{Fossorier01}
M.~P.~C. Fossorier, ``Iterative reliability-based decoding of low-density
  parity check codes,'' \emph{IEEE Transactions on Information Theory}, pp.
  908--917, May 2001.

\bibitem{PishroFekri}
H.~Pishro-Nik and F.~Fekri, ``On decoding of {LDPC} codes over the erasure
  channel,'' \emph{IEEE Trans. Inform. Theory}, vol.~50, pp. 439--454, 2004.

\bibitem{YanFelWan06}
K.~Yang, J.~Feldman, and X.~Wang, ``Nonlinear programming approaches to
  decoding low-density parity-check codes,'' \emph{IEEE J. Sel. Areas in
  Communication}, vol.~24, no.~8, pp. 1603--1613, August 2006.

\bibitem{CheChe06}
M.~Chertkov and V.~Y. Chernyak, ``Loop calculus helps to improve belief
  propagation and linear programming decoding of ldpc codes,'' in
  \emph{Allerton Conference on Communications, Control and Computing},
  Monticello, IL, September 2006.

\bibitem{Tanner80}
R.~M. Tanner, ``A recursive approach to low complexity codes,'' \emph{IEEE
  Trans. Info. Theory}, vol. IT-27, pp. 533--547, September 1980.

\bibitem{Frank01}
F.~Kschischang, B.~Frey, and H.-A. Loeliger, ``Factor graphs and the
  sum-product algorithm,'' \emph{IEEE Trans. Info. Theory}, vol.~47, no.~2, pp.
  498--519, February 2001.

\bibitem{Berlekamp78}
E.~Berlekamp, R.~McEliece, and H.~van Tilborg, ``On the inherent intractability
  of certain coding problems,'' \emph{IEEE Trans. Info. Theory}, pp. 384--386,
  1978.

\bibitem{WaiJor03Monster}
M.~J. Wainwright and M.~I. Jordan, ``Graphical models, exponential families,
  and variational inference,'' UC Berkeley, Department of Statistics, No. 649,
  Tech. Rep., September 2003.

\bibitem{SipSpi96}
M.~Sipser and D.~Spielman, ``Expander codes,'' \emph{IEEE Trans. Info. Theory},
  vol.~42, pp. 1710--1722, November 1996.

\bibitem{bargzemor}
A.~Barg and G.~Z\'{e}mor, ``Error exponents of expander codes,'' \emph{IEEE
  Trans. on Information Theory}, vol.~48, no.~6, pp. 1725--1729, 2002.

\bibitem{Ziegler}
G.~M. Ziegler, \emph{Lectures on polytopes}.\hskip 1em plus 0.5em minus
  0.4em\relax New York: Springer-Verlag, 1995.

\end{thebibliography}

\end{document}